\documentclass[twocolumn]{aastex63_cranmer2021}
\usepackage{times}

\shorttitle{Inward Propagating Plasma Parcels in the Solar Corona}
\shortauthors{Cranmer, DeForest, \& Gibson}

\begin{document}

\title{Inward Propagating Plasma Parcels in the Solar Corona:
Models with Aerodynamic Drag, Ablation, and \\ Snowplow Accretion}

\correspondingauthor{Steven R. Cranmer}
\author[0000-0002-3699-3134]{Steven R. Cranmer}
\affiliation{Department of Astrophysical and Planetary Sciences,
Laboratory for Atmospheric and Space Physics,
University of Colorado, Boulder, CO 80309, USA}

\author[0000-0002-7164-2786]{Craig E. DeForest}
\affiliation{Southwest Research Institute,
1050 Walnut Street, Suite 300, Boulder, CO 80302, USA}

\author[0000-0001-9831-2640]{Sarah E. Gibson}
\affiliation{National Center for Atmospheric Research,
3080 Center Green Drive, Boulder, CO 80301, USA}

\begin{abstract}
Although the solar wind flows primarily outward from the Sun to
interplanetary space, there are times when small-scale plasma inflows
are observed.
Inward-propagating density fluctuations in polar coronal holes were
detected by the COR2 coronagraph on board the STEREO-A spacecraft
at heliocentric distances of 7 to 12 solar radii, and these
fluctuations appear to undergo substantial deceleration as they move
closer to the Sun.
Models of linear magnetohydrodynamic waves have not been able to
explain these deceleration patterns, so they have been interpreted more
recently as jets from coronal sites of magnetic reconnection.
In this paper, we develop a range of dynamical models of discrete
plasma parcels with the goal of better understanding the observed
deceleration trend.
We found that parcels with a constant mass do not behave like the
observed flows, and neither do parcels undergoing ablative mass loss.
However, parcels that accrete mass in a snowplow-like fashion can
become decelerated as observed.
We also extrapolated OMNI in~situ data down to the so-called
Alfv\'{e}n surface and found that the initial launch-point for the
observed parcels may often be above this critical radius.
In other words, in order for the parcels to flow back down to the Sun,
their initial speeds are probably somewhat nonlinear (i.e.,
supra-Alfv\'{e}nic) and thus the parcels may be associated with
structures such as shocks, jets, or shear instabilities.
\end{abstract}

\keywords{Astrophysical fluid dynamics (101) --
Magnetohydrodynamics (1964) --
Solar corona (1483) --
Solar magnetic reconnection (1504) --
Solar wind (1534) --
Space plasmas (1544)}

\section{Introduction}
\label{sec:intro}

The Sun continuously releases a fraction of its own mass in the form
of an accelerating outflow of ionized gas called the solar wind.
There are many direct and indirect measurements of the outward
propagation of plasma parcels above the Sun's surface
\citep[see, e.g.,][]{Sh97,Schwenn,Ko06,To10,Ab16}.
However, we do sometimes observe plasma flows that go
back down to the solar surface.
Close to the Sun, there are supra-arcade downflows in hot
active regions \citep{Mc00,Sv12}, infalling droplets within
cool prominences \citep{HB11}, and ``coronal rain'' occurring in
loop-like structures \citep{Mu05,An15}.
Further up, at heliocentric distances of about 2--6~$R_{\odot}$
(solar radii), there are signatures of inflowing parcels in
visible-light coronagraphy \citep{Wa99,WS02,SW14,Do14,SD17}.
Inflows at even larger distances of 7--15~$R_{\odot}$ were
detected by \citet{Df14} via the careful Fourier filtering of
coronagraphic data from the COR2 instrument on the STEREO-A
spacecraft \citep{Hw08}.

It is likely that the reason no coronal inflows are observed above
15--20~$R_{\odot}$ is because this is the vicinity of the Sun's
Alfv\'{e}n radius $r_{\rm A}$.
This boundary---sometimes also called the Alfv\'{e}n surface or the
Alfv\'{e}n point--- is where the solar wind speed begins to exceed the
characteristic wave speeds of all magnetohydrodynamic (MHD) waves.
Thus, it acts as a critical surface beyond which information cannot
propagate back down to the Sun in the form of linear waves
\citep{WD67,BM76}.
It is also thought to be an effective ``source surface'' for heliospheric
magnetic flux, since field lines that extend beyond $r_{\rm A}$
appear to have no way of connecting back down to the photosphere.
MHD simulations of the corona--heliosphere system tend to place
the Alfv\'{e}n radius at distances between 10 and 30 $R_{\odot}$
\citep[e.g.,][]{Pi11,Coh15,Chh19}, and the three-dimensional
shape of the surface is often far from spherical.

The COR2 coronal-hole measurements of \citet{Df14}
indicated an overall {\em deceleration} for inflowing parcels:
from speeds of order 80 km~s$^{-1}$ at 12~$R_{\odot}$ to
speeds of 20 km~s$^{-1}$ by the time they reach 7~$R_{\odot}$.
However, models of MHD-wave propagation \citep[e.g.,][]{Tn16} were
not able to reproduce this observed trend.
Alfv\'{e}n or fast-mode waves that propagate radially back to the
Sun ought to undergo acceleration, not deceleration.
Obliquely propagating MHD waves may decelerate as they become
refracted, but they would exhibit the opposite ``concavity''
(in a radius-versus-speed diagram) than the COR2 data.\footnote{%
Illustrations of this concavity discrepancy can be found in
Figure \ref{fig03}, below, or in Figures 2--3 of \citet{Tn16}.}
Thus, \citet{Tn16} concluded that these parcels are not wavelike
oscillations, and may instead be jets from coronal sites of
magnetic reconnection.
In general, such parcels could be bursty flows associated with
localized reconnection events at the tips of streamers or pseudostreamers
\citep{Wa98,Rp12,SD19}, tearing-mode islands from more distributed
sites of turbulent reconnection \citep{Sm04,Gs07,Rv20}, or possibly
Kelvin-Helmholtz vortices driven by sheared solar-wind flows
\citep{Ro92,OT11}.

This paper explores a range of dynamical models of parcel motion
through the corona, and we attempt to identify the model parameters
that best reproduce the observed deceleration trend of \citet{Df14}.
In Section \ref{sec:drag} we describe a set of equations of motion
inspired by drag-based studies of coronal mass ejections (CMEs).
Section \ref{sec:snowplow} generalizes the model to allow for
parcels to accrete mass from the ambient solar wind plasma, and
Section \ref{sec:ablate} includes the effects of ablative mass loss
from the parcels.
Although there are some astrophysical environments where mass gain
and mass loss can compete with one another (e.g., cold clouds in the
circumgalactic medium; see \citeauthor{GO18} \citeyear{GO18};
\citeauthor{Li20} \citeyear{Li20}), here we keep the two models
separate and distinct.
In Section \ref{sec:obscon} we describe observational constraints on
the initial conditions of the parcel at 12~$R_{\odot}$, and we use OMNI
in~situ data to estimate probability distributions for characteristic
inflow speeds at this radius.
Notably, we discuss the possibility that downward flows can exist
{\em above} the Alfv\'{e}n surface if the initial conditions are
nonlinear (i.e., locally supra-Alfv\'{e}nic).
Lastly, Section \ref{sec:conc} concludes by summarizing our results,
suggesting future improvements to these kinds of models, and
discussing some broader implications.

\section{The Aerodynamic Drag Model}
\label{sec:drag}

The kinematic interaction between the proposed parcels and the
background solar-wind plasma is described using a drag-based formalism
adapted from CME modeling (Section \ref{sec:drag:eqns}),
with both analytic (Section \ref{sec:drag:analytic}) and
numerical (Section \ref{sec:drag:numerical}) solutions
explored below.
In reality, the physical origin of the parcel---e.g.,
magnetic reconnection, shock steepening, or MHD instability---will
likely have something to do with its subsequent dynamics.
However, here we take a slightly agnostic approach and consider
general forces that can apply in many possible scenarios.

\subsection{Governing Equations}
\label{sec:drag:eqns}

The complete equation of motion for a small coronal parcel would
require the inclusion of forces from gravity, the gas-pressure
gradient, the wave-pressure gradient \citep{BG68,B71},
magnetic mirror effects (if the velocity distribution is
anisotropic with respect to the magnetic field),
MHD Lorentz-force terms, and particle collisions.
For both the parcel and the surrounding solar wind,
many of these forces are expected to be comparable to one another
in magnitude.
Thus, as in studies of CME kinematics \citep[e.g.,][]{Cg04},
we will account only for relative differences in forcing between
the parcel and its surroundings.
By specifying the acceleration of the ambient solar wind as a kind
of attractor for the parcel flow, we can neglect the forces
that maintain it in its assumed steady-state.

Hydrodynamic and MHD forces of interaction between the parcel
and the surrounding solar wind are often referred to as
{\em drag forces.}
When studying the motions of CMEs, there has been some debate about
whether this force scales linearly with the relative velocity
between the two fluids (as in high-viscosity Stokes drag; see
\citeauthor{Vr01} \citeyear{Vr01}), or whether it scales
quadratically (as in low-viscosity turbulent drag; see
\citeauthor{Bg09} \citeyear{Bg09}).
Comparison with MHD simulations tends to favor quadratic drag
\citep[e.g.,][]{Cg96,Cg04}, so that formalism is used here.
Consider a finite parcel with mass $M_i$, positioned at a
heliocentric distance $r_i$ and flowing radially with velocity $u_i$.
Its equations of motion are
\begin{equation}
  \frac{dr_i}{dt} \,\, = \,\, u_i \,\, ,
  \label{eq:eomri}
\end{equation}
\begin{equation}
  (M_i + M_{\rm add}) \frac{du_i}{dt} \,\, = \,\, -\rho_e A C_{\rm D}
  (u_i - u_e) | u_i - u_e |
  \label{eq:eomui}
\end{equation}
where subscript $i$ refers to the internal properties of the parcel
and subscript $e$ refers to the external properties of the
surrounding corona
\citep[see also][]{Ch89,Ch96,Sub12,Vr13,Do14,Du18,Kay20}.
Thus, $u_e$ and $\rho_e$ are specified as the speed and mass density
of the ambient, time-steady solar wind.
From the form of Equation (\ref{eq:eomui}), it is clear that
the asymptotic solution is for $u_i$ to approach $u_e$ in the
limit of $t \rightarrow \infty$.

\begin{figure}[!t]
\epsscale{1.05}
\plotone{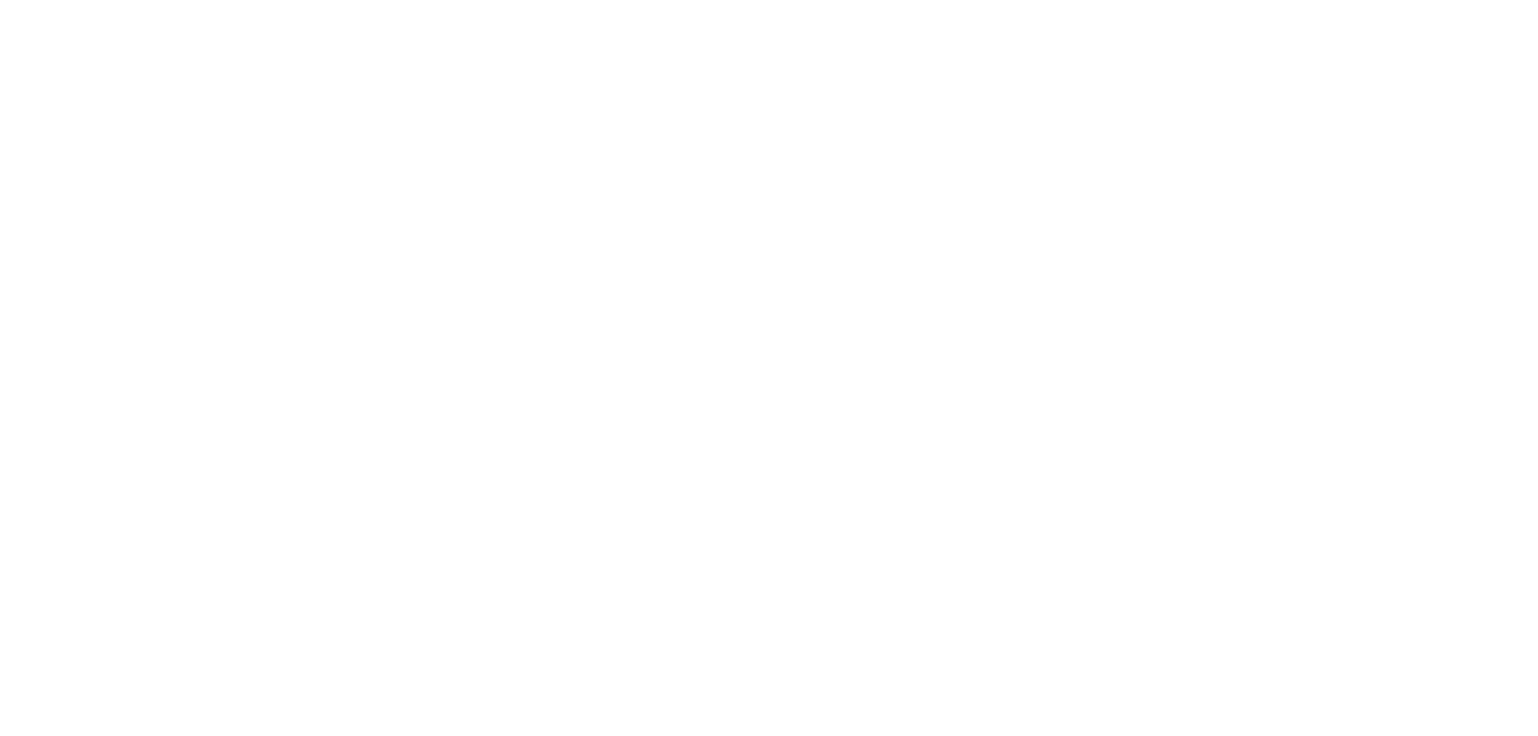}
\caption{Simplified illustration of a solar-wind plasma parcel
moving radially along a superradially expanding magnetic flux tube.
See text for definitions of the parcel parameters.
Light gray arrows indicate possible streamlines of the
surrounding flow (relative to the parcel).
\label{fig01}}
\end{figure}

Figure \ref{fig01} illustrates a simplified parcel geometry, in which
the volume is given by the product of a radial length-scale $L$ and a
cross-sectional area $A$.
The mass interior to the parcel is written as
\begin{equation}
  M_i \,\, = \,\, \rho_i A L \,\, ,
\end{equation}
where all three components---including the parcel's mean mass
density $\rho_i$---can be allowed to vary with radial distance.
For the models discussed in this section, it is assumed that
$M_i$ remains constant in time.
Note that the left-hand side of Equation (\ref{eq:eomui}) also
contains an added mass (or virtual mass) term
\begin{equation}
  M_{\rm add} \,\, = \,\, C_{\rm A} M_e
\end{equation}
where $M_e = \rho_e A L$ is the mass of adjacent solar-wind plasma
with the same volume as the moving parcel.
The dimensionless coefficient $C_{\rm A}$ is often assumed
to be equal to 1/2, and it is discussed in more detail below.
The $M_{\rm add}$ term accounts for the fact that a finite-sized
parcel induces motions in the surrounding plasma, and thus carries
along a bit of extra inertia.
For additional applications of this concept to solar flux tubes,
see \citet{Sp81}, \citet{RP93}, and \citet{CvB05}.

Much of the physics describing momentum transfer between the parcel
and its surroundings is encapsulated in the dimensionless
drag coefficient $C_{\rm D}$.
Computational and observational studies of CMEs have tried to use
the large-scale motions of flux ropes to determine likely values
for $C_{\rm D}$ in the corona and inner heliosphere.
For example, \citet{Cg96} found that varying the magnetic-field
geometry in MHD simulations caused $C_{\rm D}$ to vary between
values of 0 and 3.
In this paper, where parcels are tracked primarily between about
7 and 12 $R_{\odot}$ (i.e., only over a relatively small dynamic range
in distance), $C_{\rm D}$ is treated as an arbitrary constant.
In situations where the parcel is interpreted as a compact MHD
plasmoid, an alternate approach could be to take account of the
full set of diamagnetic forces, including a ``melon-seed'' effect
that accelerates the object in the direction of the weaker external
field \citep[see, e.g.,][]{Schluter,P57,PC85,Md90,Lin08}.

Realistic values for the added-mass coefficient $C_{\rm A}$ can be
obtained from hydrodynamics.
The standard value of 1/2, used in many drag-based CME models, is
appropriate for a spherical object embedded in an incompresible fluid.
However, the parcels considered here may become highly anisotropic;
i.e., either prolate (longer in the radial direction than in the
transverse direction) or oblate (shorter in the radial direction
than in the transverse direction).
Thus, to take one step beyond the assumption of sphericity, we
utilize results for {\em spheroidal objects} in incompressible flows
\citep[e.g.,][]{Lamb,Mo15,Fz17}.
For a spheroid moving along the direction of its primary axis
of symmetry, the total length along its trajectory is specified as
$L_{\parallel}$, and its transverse diameter (i.e.,
the diameter of the circle one would see by looking down its axis)
is $L_{\perp}$.
Defining the aspect ratio $\alpha = L_{\parallel}/L_{\perp}$, the
spheroid is oblate if $\alpha < 1$ and prolate if $\alpha > 1$.

\begin{figure}[!t]
\epsscale{1.15}
\plotone{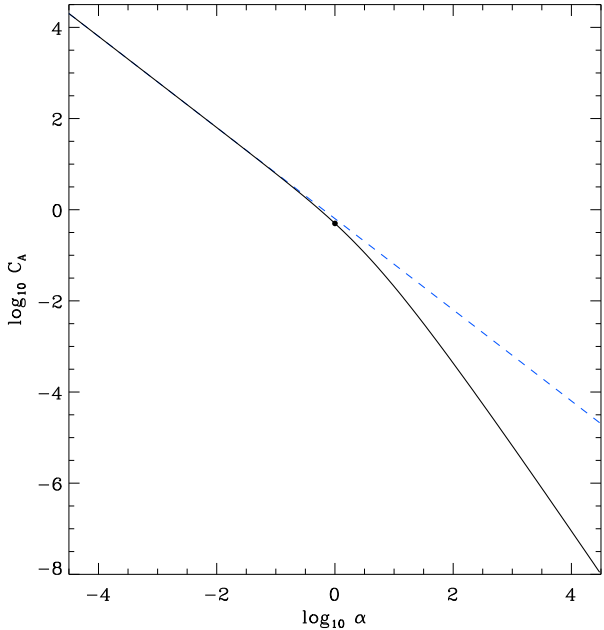}
\caption{Added mass coefficient $C_{\rm A}$ plotted versus
spheroidal aspect ratio $\alpha$ (solid black curve),
computed from formulas given by \citet{Fz17}.
Also shown is the $\alpha \ll 1$ limit (dashed blue curve)
and the spherical case of $\alpha = 1$ and $C_{\rm A} = 1/2$
(black circle).
If the fitting formula given in Equation (\ref{eq:CAfit}) were
overplotted, it would not be distinguishable from the exact solution.
\label{fig02}}
\end{figure}

Figure \ref{fig02} shows the result of an ideal hydrodynamics
calculation for $C_{\rm A}$ as a function of $\alpha$.
The exact functions are lengthy, but are given in full in
Equations 7.139--7.140 and 7.162--7.163 of \citet{Fz17}.
In the limit of $\alpha \ll 1$, the analytic result reduces to
$C_{\rm A} \approx 2/(\pi \alpha)$.
Note that $C_{\rm A} \rightarrow \infty$ in this case because the
parcel becomes equivalent to a flat disk whose volume approaches zero.
The added mass $M_{\rm add}$ remains finite and is 
given by $2 \rho_e / \pi$ times the volume of an equivalent
sphere that circumscribes the disk exactly.
In the limit of $\alpha \gg 1$, the analytic result reduces to
$C_{\rm A} \approx \ln (4\alpha^2) / (2\alpha^2)$,
which approaches zero as $\alpha \rightarrow \infty$.
For the parameters shown in Figure \ref{fig02}, the following
fitting formula always agrees with the exact calculation to within 5\%,
\begin{equation}
  C_{\rm A} \,\, \approx \,\, \frac{2}{\pi \alpha
  (1 + 0.287105 \alpha^{0.843474})} \,\, ,
  \label{eq:CAfit}
\end{equation}
and this expression is used in the numerical code that solves
for the parcel's evolution, with $L = L_{\parallel}$
and $A = \pi L_{\perp}^2 / 4$.

In order to determine all parameters in the model, there still needs
to be a specification of the parcel's time-dependent shape and
anisotropy.
The cross-sectional area $A$ is assumed to be identical to that of
the large-scale background magnetic field; i.e., we assume
that the parcel expands or contracts laterally to fill its local
magnetic flux tube.
As long as the radial magnetic-field strength $B_r$ is known, the
principle of magnetic flux conservation provides the relative radial
variation of $A \propto B_r^{-1}$.
Thus, a given model of the solar wind has unique values for
$u_e(r)$, $\rho_e(r)$, and $A(r)$.
The evolution in radial length of the parcel is specified with a
scaling exponent $\sigma$, as
\begin{equation}
  L \,\, = \,\, L_0 \left( \frac{A}{A_0} \right)^{\sigma}
  \label{eq:Lshape}
\end{equation}
where quantities with subscript 0 are the initial conditions at $t=0$.
The exponent $\sigma$ will be varied as a free parameter, and it
is useful to note that there are several natural values that may
pertain in different circumstances:
\begin{enumerate}
\item
A value of $\sigma = -1$ would be appropriate if the volume of
the parcel remains constant over time.
This may occur for a parcel that retains a constant mass
($M_i =$~constant) and is close to incompressible
($\rho_i =$~constant).
\item
A value of $\sigma = 0$ constrains the length $L$ to remain constant.
This may be the case if the forces on the front and back faces of
the parcel always remain roughly equal to one another.
It may also be occurring in the outer corona for observed
visible-light ``flocculations'' that appear to preserve their
radial length-scale from about 30 to 90 $R_{\odot}$
\citep[see Figure 9 of][]{Df16}.
\item
The specific value $\sigma \approx 0.14$ would occur if the
parcel's mass is conserved ($M_i =$~constant) and its internal
density remains proportional to the surrounding solar-wind density
($\rho_i \propto \rho_e$).
This value comes from the ZEPHYR coronal-hole model discussed
below, in which $\rho_e \propto r^{-2.45}$ and
$A \propto r^{2.15}$ over the relevant range of heights to be
compared with the observational data (7--12~$R_{\odot}$).
\item
Lastly, the value $\sigma = 0.5$ would occur if the aspect ratio
$\alpha$ of the parcel remains constant over time.
An initially spherical parcel may remain so if the
local pressure-gradient forces on it provide a roughly
isotropic confinement.
\end{enumerate}
It turns out that the models that successfully explain the observed
COR2 parcel deceleration do not depend sensitively on
the exponent $\sigma$.
Thus, treating it as a free parameter probably does not limit
the usefulness of these models.
For another approach to simulating the evolution of parcel
prolateness in the accelerating solar wind, see \citet{PC85}.

For a specified background solar wind, the above model has two
free parameters ($C_{\rm D}$ and $\sigma$) and five initial
conditions ($r_{i,0}$, $u_{i,0}$, $M_{i,0}$, $L_0$, and $A_0$).
The properties that are specified explicitly are
$r_{i,0}$, $u_{i,0}$, $L_0$, and the initial value of
the density ratio $\rho_i/\rho_e$.
The initial parcel shape is assumed to be spherical, with
$A_0 = \pi L_0^2 / 4$, and $M_{i,0}$ is computed
from the other quantities.
For now, the parcel mass is presumed to remain constant over
time, with $M_{i}(t) = M_{i,0}$.

\subsection{Approximate Analytic Solutions}
\label{sec:drag:analytic}

Equations (\ref{eq:eomri}) and (\ref{eq:eomui}) will eventually be
solved numerically, but it is also helpful to explore closed-form
solutions when possible.
Thus, we make the simplifying assumption that the wind speed
$u_e$, background density $\rho_e$, and cross-sectional area $A$
are all constants as a function of radial distance.
This also constrains $\rho_i$ and $L$ to be constants as well.
The assumption of spherical symmetry for the parcel gives the
standard hydrodynamic value for the added-mass coefficient,
$C_{\rm A} = 1/2$ \citep{Lamb}.

Because essentially nothing in the simplified system depends on
radial distance (other than the primary dependent variable $u_i$),
Equation (\ref{eq:eomui}) can be solved first for $u_i(t)$, then
the radial trajectory $r_i(t)$ can be determined subsequently.
It is convenient to define two new dependent variables
\begin{equation}
  U \, = \, | u_i - u_e |
  \,\,\,\,\,\,\,\,\, \mbox{and}
  \,\,\,\,\,\,\,\,\,
  Q \, = \, \frac{\rho_i}{\rho_e} + \frac{1}{2} \,\, ,
  \label{eq:UQdef}
\end{equation}
where the two terms in $Q$ correspond to the $M_i$ and $M_{\rm add}$
terms in Equation (\ref{eq:eomui}).
For now, the assumption is that $Q$ remains fixed at its initial
value $Q_0$.
Equation (\ref{eq:eomui}) is thus rewritten as
\begin{equation}
  \frac{dU}{dt} \, = \, - \frac{C_{\rm D} \, U^2}{Q_0 L} \,\, .
  \label{eq:andrag}
\end{equation}
If all quantities on the right-hand side are positive, then
$U$ will get smaller as time increases---i.e., asymptotically
approaching zero as the parcel's motion becomes entrained
into the surrounding solar wind.
However, if the initial condition obeys $u_i < u_e$, then the
time evolution of $u_i$ can go from negative (inward flow)
to positive (entrained outward flow).
For a known initial condition $U_0$, this equation has an
explicit solution
\begin{equation}
  U(t) \,\, = \,\, U_0 (1 + \omega_0 t)^{-1}
  \label{eq:Umasscon}
\end{equation}
with
\begin{equation}
  \omega_0 \, = \, \frac{C_{\rm D} \, U_0}{Q_0 L} \,\, .
  \label{eq:omega0}
\end{equation}
Note that significant deceleration occurs with a characteristic
time scale of $t \sim 1 / \omega_0$.
The radial position of the parcel is given by integrating
Equation (\ref{eq:eomri}),
\begin{equation}
  r_i (t) \, = \, r_{i,0} + u_e t - \frac{U_0}{\omega_0}
  \ln ( 1 + \omega_0 t )
  \label{eq:Rmasscon}
\end{equation}
where a sign convention was adopted that assumes $u_i < u_e$ for
all $t \geq 0$.

\begin{figure}[!t]
\epsscale{1.15}
\plotone{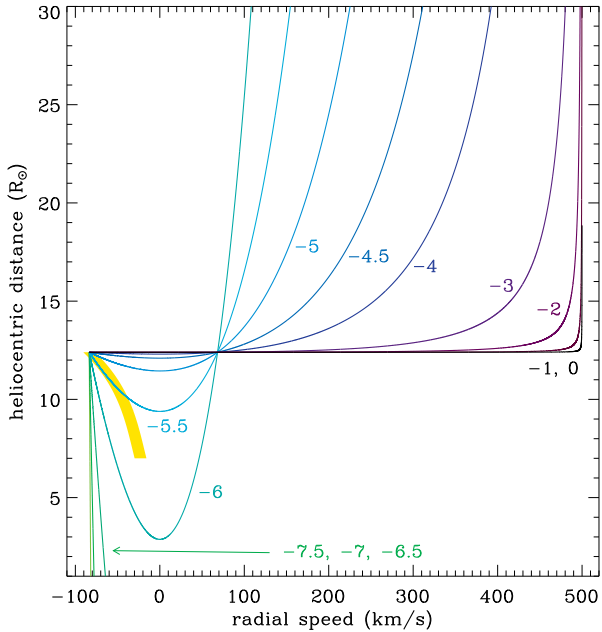}
\caption{Example analytic solutions for a parcel's radial distance
$r_i$ versus its instantaneous radial velocity $u_i$, modeled with
no mass gain or loss (i.e., $M_i =$~constant).
Solid curves are labeled by their values of $\log_{10} \omega_0$,
where $\omega_0$ is expressed in s$^{-1}$.
Also shown (yellow region) is the observed COR2 inflow ridge from
\citet{Df14}.
\label{fig03}}
\end{figure}

Figure \ref{fig03} shows a range of illustrative solutions of
Equations (\ref{eq:Umasscon}) and (\ref{eq:Rmasscon}).
The initial conditions were chosen to agree with the observations
of \citet{Df14} for the largest height at which inflowing parcels
were detected: $r_{i,0} = 12.4 \, R_{\odot}$ and
$u_{i,0} = -83.6$ km~s$^{-1}$.
Here, we also chose a fast solar wind speed of
$u_e = 500$ km~s$^{-1}$, typical for coronal holes at this distance.

The only other parameter required to implement the above solutions
is $\omega_0$, and we can estimate likely values by examining the
components of Equation (\ref{eq:omega0}).
From the CME literature, a realistic range of values for
$C_{\rm D}$ seems to be 0.001--1.
The modified overdensity ratio $Q_0$ may be as small as 0.5
(if $\rho_i < \rho_e$) and possibly as high as 100.
The observations of \citet{Df14} found that the parcel length-scale
$L$ is probably no larger than about 0.5~$R_{\odot}$.
Although it is unclear what the smallest value of $L$ could be,
the fact that these features are observable with STEREO/COR2
probably means they are not much smaller than one pixel in size.
Thus, we adopt a single-pixel lower limit for $L$ of about 15$''$,
or 0.016~$R_{\odot}$ \citep{Hw08}.
Combining the ranges for $C_{\rm D}$, $Q_0$, and $L$ with the initial
speeds discussed above (i.e., $U_0 = 583.6$ km~s$^{-1}$) gives a
possible spread of about eight orders of magnitude for $\omega_0$.
Figure \ref{fig03} shows the dynamical solutions for values between
$3 \times 10^{-8}$ s$^{-1}$ (green curve) and
1 s$^{-1}$ (black curve).

The green curves in Figure \ref{fig03} correspond to very low values
of $\omega_0$.
In this case, the drag force on the parcels is extremely weak;
they just continue to flow at their initial downward velocity
and eventually reach the solar surface.
The violet and black curves in Figure \ref{fig03} correspond to
extremely large values of $\omega_0$.
In that case, the parcels become entrained very quickly into the ambient
solar wind and flow out with speeds approaching $u_e = 500$ km~s$^{-1}$.
Note that all of the entrained solutions pass through the
the initial radius $r_{i,0}$ for a second time at the same velocity,
$u_i = 68.4$ km~s$^{-1}$.
It is possible to write an analytic solution for this velocity, but it
requires the use of the Lambert $W$ function \citep{Cor96}, with
\begin{equation}
  u_i \, = \, u_e \left\{ 1 + \left[
  W_{-1} \left( -\frac{u_e}{U_0} e^{-u_e / U_0} \right) \right]^{-1}
  \right\}  \,\, ,
\end{equation}
and these solutions all exhibit $\omega_0 t = 0.35207$.

Figure \ref{fig03} also shows a fit to the ``inflow ridge''
(yellow region) that indicates the observed trend of parcel
deceleration \citep{Df14} and is given by
\begin{equation}
  u_{i,{\rm{fit}}} \, = \,
  118.39 - 46.739 r_i + 5.5295 r_i^2 - 0.24790 r_i^3
\end{equation}
where $r_i$ is in units of $R_{\odot}$ and
$u_{i,{\rm{fit}}}$ is in units of km~s$^{-1}$.
This fitting formula is valid only between 7 and 12.4 $R_{\odot}$
and should not be applied outside this range.
Note, however, that no choice of $\omega_0$ matches the data well.
Like the linear MHD-wave modes studied by \citet{Tn16}, the model
curves have the opposite concavity as the observed trend.

\subsection{Numerical Solutions}
\label{sec:drag:numerical}

\begin{figure}[!t]
\epsscale{1.15}
\plotone{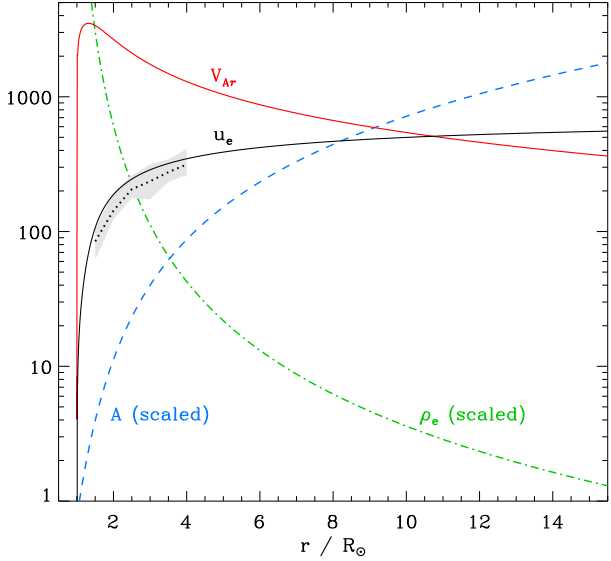}
\caption{Radial variation of several coronal-hole parameters from
the ZEPHYR model, including the wind speed $u_e$ (black solid curve)
and Alfv\'{e}n speed $V_{{\rm A}r}$ (red solid curve), both given in
units of km~s$^{-1}$.
Also shown are scaled (i.e., arbitrarily renormalized) radial trends
for the cross-sectional area $A$ (blue dashed curve) and the
ambient density $\rho_e$ (green dot-dashed curve).
For comparison, Doppler-dimming wind-speed measurements from
UVCS/SOHO \citep{Cr20a,Cr20b} are also shown, with
median values (black dotted curve) shown with $\pm 1$ standard
deviation uncertainty bounds (gray region).
\label{fig04}}
\end{figure}

It is possible that the analytic model described
in Section \ref{sec:drag:analytic} fails to match the data
because its assumptions about constant values for the model
parameters (e.g., $u_e$, $\rho_e$, $A$, $L$) were too simplistic.
Thus, more realistic radial dependences for these plasma properties
can be adopted, and the original drag equations can be solved
numerically.
Figure \ref{fig04} shows coronal-hole parameters from ZEPHYR, a
one-fluid model of turbulent heating and acceleration along
open magnetic flux tubes \citep{CvB07}.
Along with $u_e(r)$, $A(r)$, and $\rho_e(r)$, we also show the
radial Alfv\'{e}n speed
\begin{equation}
  V_{{\rm A}r} \, = \, \frac{B_r}{\sqrt{4\pi \rho_e}}
  \label{eq:VA}
\end{equation}
in the vicinity of relevant radii for the COR2 parcels.
For this model, the Alfv\'{e}n radius is located at
$r_{\rm A} = 10.73 \, R_{\odot}$, which is notably {\em below}
the initial-condition parcel radius of 12.4~$R_{\odot}$.
This means that a parcel that flows back down to the Sun from
12.4~$R_{\odot}$ must have an initial speed that is locally
faster than the Alfv\'{e}n speed.
Some implications of this situation are discussed in
Section \ref{sec:obscon}, but for now we will assume that
parcels can flow back toward the Sun at all radii
$\leq 12.4 \, R_{\odot}$.

A numerical code was developed to integrate
Equations (\ref{eq:eomri}) and (\ref{eq:eomui}) forward in time
using first-order Euler steps of size
$\Delta t = 10^{-4} R_{\odot}/|u_{i,0}|$.
At each step, the current value of $r_i$ was used to interpolate
the local values of $u_e$, $\rho_e$, and $A$ from the ZEPHYR model
grids, and to compute the local values of $L$ and $\rho_i$ from
the expressions given in Section \ref{sec:drag:eqns}.
At least 10,000 trial models were constructed with randomized
choices for four key parameters:
(1) $C_{\rm D}$ was sampled from a uniform grid (in logarithm)
between $10^{-6}$ and $10^{+1}$;
(2) the shape exponent $\sigma$ was sampled from a uniform grid
of values between $-6$ and $+6$;
(3) the initial parcel length $L_0$ was sampled from a uniform
grid (in logarithm) between 0.001~$R_{\odot}$ and 0.5~$R_{\odot}$;
(4) the initial parcel overdensity ratio $(\rho_i / \rho_e)_0$
was sampled from a uniform grid (in logarithm) between
$10^{-3}$ and $10^{+3}$.

For each numerical model, the values of $u_i$ at three specified
radial distances were saved, compared with the observed inflow speeds,
and used to compute a reduced $\chi^2$ goodness-of-fit parameter,
\begin{equation}
  \chi^2 \, = \, \frac{1}{N} \sum_{j=1}^{N} \left[
  \frac{u_i (r_j) - u_{i,{\rm obs}} (r_j)}{\delta u_i} \right]^2
\end{equation}
where $N=3$.
The sum is taken over the observations at
$r_j = \{ 7, 9, 11 \} R_{\odot}$, with corresponding velocities
$u_{i,{\rm obs}} = \{ -22.86, -35.08, -56.62 \}$ km~s$^{-1}$.
A somewhat arbitrary uncertainty width of $\delta u_i = 5$ km~s$^{-1}$
was adopted as well; see Figure~7 of \citet{Df14}.
If a given trial model never made it down to 7~$R_{\odot}$ (i.e.,
if it was rapidly entrained into the outflowing solar wind like
the high-$\omega_0$ models in Figure \ref{fig03}), it was assigned
an artificially large value of $\chi^2$ as a flag to neglect that
set of input parameters.

After randomly sampling the four-dimensional parameter space
discussed above, we found no deceleration trajectories
that ended up matching the observed trend very well.
The ``best'' models (which exhibited a minimum $\chi^2 \approx 6$)
were very similar in appearance to the analytic solutions
shown in Figure \ref{fig03} for values of $\log \omega_0$
between about $-6$ and $-5.5$.
The model with the lowest $\chi^2$ had a value of
$\omega_0 \approx 1.7 \times 10^{-6}$ s$^{-1}$, and the parcel
reached a minimum radius of about 6.5~$R_{\odot}$ before
turning around and flowing out with the solar wind.
Thus, even when taking into account the radial variation of
background solar-wind parameters (which the analytic model above
did not do) it seems that the assumption of a constant-$M_i$ parcel
cannot explain the inflow pattern observed by \citet{Df14}.

In order to better match the observed deceleration trend, it is
clear that $\omega_0$ should be allowed to vary even more than it
does in the numerical models discussed in this section.
Figure \ref{fig03} indicates that $\omega_0$ should be rather
large when the parcel is first released, then become smaller
as the parcel reaches the lowest observed height.
In other words, the magnitude of the drag acceleration
$du_i/dt$ must decrease over time.

\section{The Drag-Based Snowplow Model}
\label{sec:snowplow}

One obvious improvement that can be made to the drag-based model is to
allow the parcel mass ($M_i$) to vary as it flows through the corona.
This section explores the idea that the parcel can accrete mass
from solar-wind plasma ahead of it.
Similar mass-gain effects have been reported for CMEs
\citep[e.g.,][]{Wb96,Tp06,Df13,Fg15}, but it is still not certain
whether this straightforward ``snowplow'' effect is the most likely
explanation for the observations \citep{HV18}.
In any case, \citet{Tn16} found that a parcel undergoing a gradual
increase in density may be able to explain the inward
deceleration seen by \citet{Df14}.
Section \ref{sec:snowplow:eqns} describes the differential
equations for this snowplow model,
Section \ref{sec:snowplow:analytic} gives analytic solutions for
simplified background properties, and
Section \ref{sec:snowplow:numerical} explores numerical solutions.

\subsection{Governing Equations}
\label{sec:snowplow:eqns}

This model now consists of three coupled differential equations
that must be solved simultaneously:
Equations (\ref{eq:eomri}), (\ref{eq:eomui}), and a new equation
that describes the rate of change of parcel mass.
We generalize from earlier versions of this kind of snowplow
(or bulldozer) equation, such as that of \citet{Tp06}, with
\begin{equation}
  \frac{dM_i}{dt} \,\, = \,\, \rho_e A C_{\rm S} | u_i - u_e |
  \label{eq:eomMi}
\end{equation}
\citep[see also][]{Ht07,Fg15,TS17}.
The constant $C_{\rm S}$ is introduced as an effective snowplow
efficiency that describes what fraction of the plasma in front of
the parcel is actually incorporated into it.
Note that \citet{Tp06} assumed $C_{\rm S} = 1$, which describes the
parcel gaining an incremental mass $dM_i$ as it traverses a radial
length $dL = |u_i - u_e| dt$ in the rest frame of the fluid, thus
sweeping out an incremental volume $dV = A dL$ containing particles
with density $\rho_e$.
Other values of $C_{\rm S} < 1$ were considered by \citet{TS17}.
The right-hand side of Equation (\ref{eq:eomMi}) is always positive,
so $M_i$ grows monotonically over time.
Thus, Equation (\ref{eq:eomui}) says that the magnitude of the
acceleration $du_i/dt$ decreases over time, at least in comparison
to the constant-$M_i$ case.

\subsection{Approximate Analytic Solutions}
\label{sec:snowplow:analytic}

As in Section \ref{sec:drag:analytic}, it is possible to first
explore an analytic solution of the system in question (for the
limiting case of constant values of $u_e$, $\rho_e$, $A$, and $L$)
before solving the full equations numerically.
In this case there are two coupled differential equations to solve;
Equations (\ref{eq:eomui}) and (\ref{eq:eomMi}).
The same substitution of variables from Equation (\ref{eq:UQdef})
allows the two equations to be written as
\begin{equation}
  \frac{dU}{dt} \, = \, - \frac{C_{\rm D} \, U^2}{QL}
  \,\,\,\,\,\,\,\,\, \mbox{and}
  \,\,\,\,\,\,\,\,\,
  \frac{dQ}{dt} \, = \, \frac{C_{\rm S} \, U}{L} \,\, .
\end{equation}
As before, $U$ gets smaller as time increases, and now the parcel-mass
proxy variable $Q$ grows larger as time increases.
The two coupled first-order equations can be combined into a single
second-order equation,
\begin{equation}
  U \, \frac{d^2 U}{dt^2} \,\, = \,\,
  \left( \frac{C_{\rm S}}{C_{\rm D}} + 2 \right)
  \left( \frac{dU}{dt} \right)^2
\end{equation}
which has a closed-form solution of the form
\begin{equation}
  U(t) \,\, = \,\, U_0 (1 + \omega t)^{-1/\gamma}
  \label{eq:UsnowAN}
\end{equation}
with
\begin{equation}
  \gamma \, = \, \frac{C_{\rm S}}{C_{\rm D}} + 1
  \,\,\,\,\,\,\,\,\, \mbox{and}
  \,\,\,\,\,\,\,\,\,
  \omega \, = \, \frac{\gamma C_{\rm D} \, U_0}{Q_0 L} \,\, .
\end{equation}
The full solution is completed by
\begin{equation}
  Q(t) \,\, = \,\, Q_0 (1 + \omega t)^{(\gamma - 1)/\gamma} \,\, .
\end{equation}
Note that both initial conditions $U_0$ and $Q_0$ are positive,
and $\gamma \geq 1$.
The radial position of the parcel is given by integrating
Equation (\ref{eq:eomri}),
\begin{equation}
  r_i (t) \, = \, r_{i,0} + u_e t - \frac{Q_0 L}{C_{\rm S}}
  \left[ ( 1 + \omega t )^{(\gamma - 1) / \gamma} - 1 \right]
  \label{eq:RsnowAN}
\end{equation}
where the same sign convention as in Equation (\ref{eq:Rmasscon})
was used.
The solutions from Section \ref{sec:drag:analytic} are recovered
exactly in the limit of $\gamma = 1$, which is equivalent to
$C_{\rm S} = 0$ \citep[see also][]{Mj20}.

\begin{figure}[!t]
\epsscale{1.15}
\plotone{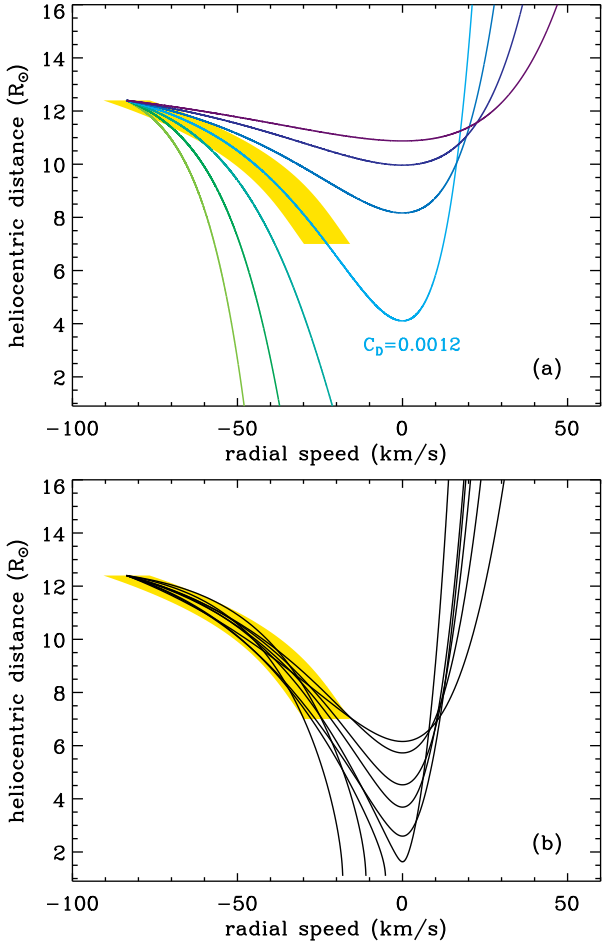}
\caption{Analytic and numerical solutions for parcel trajectories
with snowplow-like accretion.
(a) Solutions to Equations (\ref{eq:UsnowAN}) and (\ref{eq:RsnowAN})
for $C_{\rm S} = 0.03$ and a range of $C_{\rm D}$ values spaced evenly
on a logarithmic grid between a minimum of $6 \times 10^{-4}$
(left-most green curve) and a maximum of $2.4 \times 10^{-3}$
(right-most violet curve).
See text for other parameter values.
(b) Random sample of 9 numerical solutions for radially varying
background parameters, each having $\chi^2 \leq 1$ (i.e., with
parameters sampled from the histograms of Figure \ref{fig06}).
Also shown in both panels is the observed COR2 trend from
\citeauthor{Df14} (\citeyear{Df14}, yellow region).
\label{fig05}}
\end{figure}

Figure \ref{fig05}(a) shows a range of illustrative solutions.
The choices for $r_{i,0}$, $u_{i,0}$, and $u_e$ were the same
as in Section \ref{sec:drag:analytic}.
For simplicity, we also chose $Q_0 = 1.5$ (i.e., initial density
equipartition between the interior and exterior of the parcel), and
$L = 0.1 \, R_{\odot}$.
In this case, it is possible to find combinations of parameters that
match the \citet{Df14} data very well.
It was found that one specific value for the ratio of snowplow to
drag coefficients (i.e., $\gamma \approx 26$) produced excellent
agreement with the data.
The best-fitting (central) curve in Figure \ref{fig05}(a) is for a
model with $C_{\rm S} = 0.03$ and $C_{\rm D} = 0.0012$.
The other curves were computed with $C_{\rm S}$ held fixed and
$C_{\rm D}$ varied up and down by factors of (up to) two.
Lower values of $C_{\rm D}$ imply weaker drag, so an inwardly
moving parcel just keeps falling in toward the Sun.
Higher values of $C_{\rm D}$ imply stronger drag, so the parcels
are eventually swept into the outward-flowing solar wind.

For the best-fitting analytic model in Figure \ref{fig05}(a), the
parcel flows in from 12.4~$R_{\odot}$ to 7~$R_{\odot}$ over a time
of about 26.1 hours.
However, this model exhibits a value of
$\omega = 1.74 \times 10^{-4}$ s$^{-1}$, or a characteristic
deceleration timescale of only $1/\omega \approx 1.6$~hours.
This is not an inconsistency because the definition of $\omega$
involves only the initial conditions of the parcel.
If a new inverse-timescale variable was defined using the
time-variable quantities (i.e.,
$\omega' = \gamma C_{\rm D} U(t)/ Q(t) L$), the combined effects
of mass-gain and deceleration reduce its value over time.
By the time the parcel reaches 7~$R_{\odot}$, this quantity
has decreased to $\omega' \approx 10^{-5}$ s$^{-1}$, driven mainly by
an increase in $Q(t)$ from 1.5 to 23.
The corresponding timescale $1/\omega'$ is thus 27.7 hours,
comparable to the total transit time.

\subsection{Numerical Solutions}
\label{sec:snowplow:numerical}

The numerical integration code described in
Section \ref{sec:drag:numerical} was extended to solve Equations
(\ref{eq:eomri}), (\ref{eq:eomui}), and (\ref{eq:eomMi}).
These models have radially varying trends in $\rho_e$, $u_e$,
$A$, and $L$, given by the ZEPHYR polar coronal-hole model
shown in Figure \ref{fig04}.
Another large set of randomized trial models was constructed for the
parcel motion and mass-gain, this time with different seed constants
for the pseudo-random number generators than the ones used above.
The parameters $\sigma$, $L_0$, and $(\rho_i / \rho_e)_0$ were
sampled as described in Section \ref{sec:drag:numerical}, and in
this case {\em both} coefficients $C_{\rm D}$ and $C_{\rm S}$ were
sampled randomly from a uniform grid (in logarithm) between
$10^{-6}$ and $10^{+1}$.

For this set of models with parcel mass-gain, there were many solutions
that ended up matching the observed deceleration trend of \citet{Df14}.
More than 10,000 total models were generated, and the code was run
until a sample-size of 500 models with $\chi^2 \leq 1$ was accumulated.
The smallest value of $\chi^2$ in that sample was 0.059.
Figure \ref{fig05}(b) shows nine sample trajectories of solutions
with $\chi^2 \leq 1$.
Note the large variety in the possible trajectories---from parcels
that turn around just below the minimum observed height of
7~$R_{\odot}$, to ones that keep flowing monotonically down to the
solar surface---that are all reasonably consistent with the observed
velocities.

\begin{figure*}[!t]
\epsscale{1.10}
\plotone{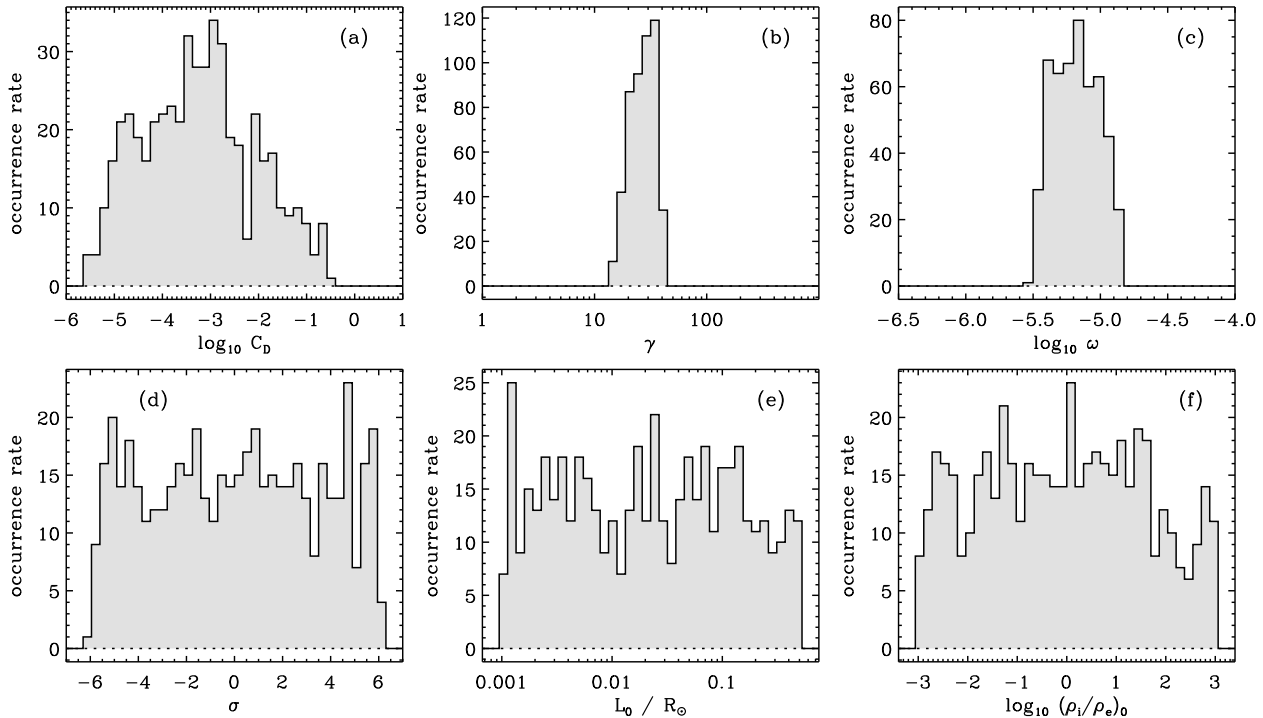}
\caption{Histograms of probability distributions for input parameters
of numerical models that agree with the \citet{Df14} inflow speeds
at a level of $\chi^2 \leq 1$.  Panels provide distributions for
(a) $C_{\rm D}$, (b) $\gamma$, (c) $\omega$, (d) $\sigma$,
(e) $L_0$, and (f) $(\rho_i / \rho_e)_0$.
\label{fig06}}
\end{figure*}

Figure \ref{fig06} shows histograms of the input parameters for the
set of 500 solutions with $\chi^2 \leq 1$.
The ranges of good-fitting values for $\gamma$ (which is essentially
the ratio of snowplow to drag coefficients) and $\omega$ (the inverse
deceleration timescale) were found to be rather narrow, with median
values of $\gamma = 27.3$ and $\omega = 6.62 \times 10^{-6}$ s$^{-1}$.
The former value is almost identical to the best-fitting value of
$\gamma$ found for the analytic models described in
Section \ref{sec:snowplow:analytic}.
For the numerical models, the drag coefficient $C_{\rm D}$ also
seemed to prefer intermediate values (with a median of
$C_{\rm D} = 5.56 \times 10^{-4}$), though the distribution of
values matching the observations was broader than in the cases of
$\gamma$ and $\omega$.
The distribution of input values for $C_{\rm S}$ is not shown, but it
had a median value of 0.0135.
The combination of (more tightly constrainted) medians for
$\gamma$ and $C_{\rm D}$ point to a slightly higher most-likely
value of $C_{\rm S} \approx 0.0146$.

It is evident from Figure \ref{fig06} that there do not seem to be any
preferred values for $\sigma$, $L_0$, or $\rho_i / \rho_e$.
Models that agree with the observed pattern of deceleration appear to
be possible for nearly any of these input parameters.
However, if we apply an independent estimate for a likely range of
values for $L_0$ (say, 0.01 to 0.1~$R_{\odot}$) it is then possible to
use the medians for $C_{\rm D}$ and $\omega$ to obtain a similarly
likely range of values for the overdensity parameter $Q_0$
between about 20 and 200.

\begin{figure}[!t]
\epsscale{1.15}
\plotone{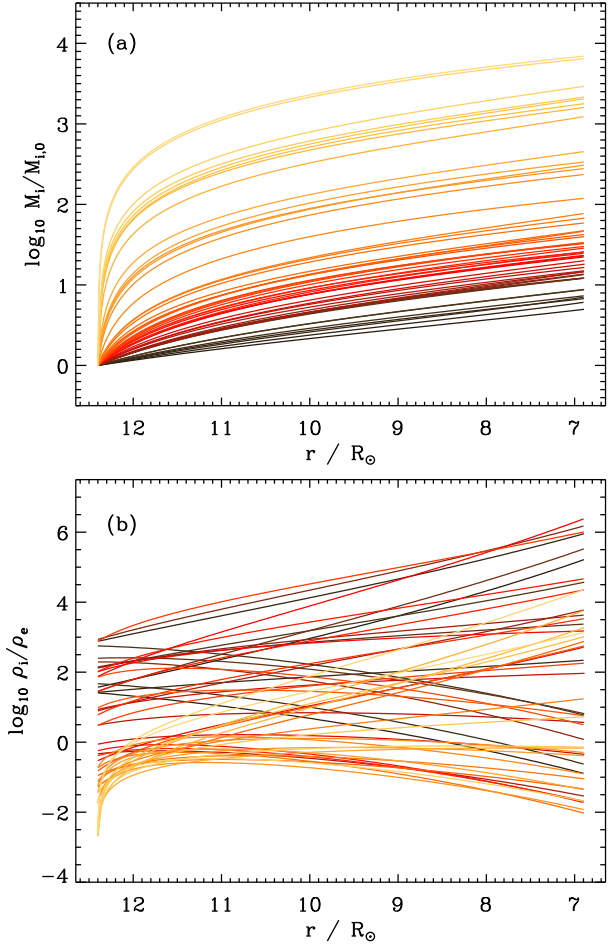}
\caption{(a) Radial dependence of the ratio of instantaneous parcel
mass $M_i$ to its initial mass $M_{i,0}$ at $r = 12.4 \, R_{\odot}$,
for 50 numerical models with $\chi^2 \leq 1$.
Curve color is proportional to the final ratio at $r = 7 \, R_{\odot}$.
(b) Radial dependence of the ratio of parcel density $\rho_i$ to
surrounding solar-wind density $\rho_e$ for the same set of models.
Each model's curve color is taken from panel (a).
\label{fig07}}
\end{figure}

Additional information about how much snowplowing occurs in these
models can be found in Figure \ref{fig07}.
Figure \ref{fig07}(a) shows the monotonic gains in parcel mass $M_i$
for a random subset of 50 models with $\chi^2 \leq 1$.
At the lowest COR2 height of 7~$R_{\odot}$, these models exhibit
a large range of final-to-initial mass ratios; i.e., between 5 and
7,000, with a median ratio of about 30.
Note that for the larger set of ``bad'' models (i.e., with
$\chi^2 > 1$; not shown) in which the parcels made it down
to 7~$R_{\odot}$, the distribution of mass-gain ratios was much
broader: sometimes as low as 1.0001, sometimes as high as 10$^5$.
Figure \ref{fig07}(b) shows how $\rho_i / \rho_e$ evolves for the
same subset of models from Figure \ref{fig07}(a).
Note that the initial range of $10^{-3}$ to $10^{+3}$ expands
gradually as the parcels evolve from 12.4 to 7~$R_{\odot}$.
Due to the broad range of randomly chosen $\sigma$ exponents (which
controls the evolution of parcel volume), there is no strong
correlation between the mass-gain ratio and the local overdensity ratio.

It remains to be determined why the best-fitting values of both
$C_{\rm D}$ and $C_{\rm S}$ fall so far below the order-unity
expectations for these dimensionless coefficients.
For the drag coefficient, it is possible that its small value
can be understood as an effect of magnetic tension.
\citet{Cg96} found that a horizontal flux rope, with an axial
field pointing along the $x$-axis and rising along the $z$-axis,
experiences strong turbulent drag with $C_{\rm D} \approx 1$.
However, if the $z$-component of the field becomes strong enough
to exceed a significant fraction of the flux rope's axial field,
the vertical magnetic tension suppresses turbulent eddy-like
vortices from forming behind the object, and reduces $C_{\rm D}$
to values near zero.
Conceptually, this may be similar to the reduction in drag when
a blunt solid object (such as a sphere) is streamlined by
lengthening and tapering its surface in the downwind direction.
In both cases, trailing streamlines become stretched out more
along the direction of motion, and the volume of the turbulent wake
behind the object is reduced drastically.
For high-Reynolds-number hydrodynamic flows, it is common to see
reductions in $C_{\rm D}$ from values of order 1 to values
of order 0.001 \citep{Hoe65,Mun09}.
Thus, a similar reduction in drag may occur for the parcels
considered here.
Although we have no firm constraints on their internal magnetic
geometry, they do seem to be moving parallel to a strong background
field as in the high-$B_z$ cases of \citet{Cg96}.

\section{The Drag-Based Ablation Model}
\label{sec:ablate}

If we want to explore models of plasma parcels with a time-evolving
mass, it is prudent to consider mechanisms of mass loss
as well as mass gain.
It has been suggested that CME flux ropes (the primary inspiration
for the snowplow models discussed above) may also undergo gradual
erosion due to magnetic reconnection at their edges
\citep{Cg96,Rf15,Pa20}.
In convectively unstable stellar interiors, there is also the
concept of a mixing-length parcel that propagates semi-coherently
until it disintegrates, thereby losing its identity and merging
into the background medium.
Mass-losing plasma ``bullets'' have also been proposed to explain
thin observed filaments in the vicinity of luminous variable stars
such as $\eta$~Car \citep{Rd02}.
Lastly, solid bodies such as meteors and planetesimals undergo
ablative mass loss as they enter the Earth's atmosphere
\citep{Op58,BS71,Br83} or are accreted by protoplanets
\citep[e.g.,][]{Md16}.
Section \ref{sec:ablate:eqns} describes the differential
equations for a proposed model of parcel ablation,
Section \ref{sec:ablate:analytic} explores analytic solutions, and
Section \ref{sec:ablate:numerical} briefly discusses numerical models.

\subsection{Governing Equations}
\label{sec:ablate:eqns}

Because the mass-gain mechanism assumed in Section \ref{sec:snowplow}
was just a local snowplow, the rate of gain was straightforwardly
proportional to the relative velocity $u_i - u_e$.
However, for ablative mass loss, it is usually assumed that the
ram pressure (i.e., kinetic energy density) of the relative flow
is the primary driver of the interaction.
Thus, the mass-loss rate is proportional to $(u_i - u_e)^3$.
We use a similar formula to that given by \citet{Br83},
\begin{equation}
  \frac{dM_i}{dt} \,\, = \,\, - \rho_e A \,
  \frac{| u_i - u_e |^3}{V_{\rm eff}^2} \, {\cal H}(M_i - M_e)
  \label{eq:dMiAblate}
\end{equation}
where $V_{\rm eff}$ is a velocity-like quantity that describes the
efficiency of the ablation.
Note the presence of a Heaviside step function ${\cal H}(M_i - M_e)$
on the right-hand side.
This constrains the ablative mass loss to occur only when
$M_i > M_e$.
The process is assumed to shut off completely when $M_i < M_e$;
i.e., when the parcel's inertia is totally dominated by that of
the surrounding solar wind.

Because there does not seem to be any well-developed theory
for the ablation of plasma parcels in the solar wind, we treat the
ablation effiency $V_{\rm eff}$ as a constant free parameter.
We also speculate about
a variety of approaches to estimating its value:
\begin{enumerate}
\item
For solid meteors entering a planetary atmosphere, $V_{\rm eff}^2$
is specified as a latent heat of ablation (with its velocity-squared
units more typically reported as energy per unit mass).
For rocky and metallic meteoric material,
$V_{\rm eff} \approx 1$--3 km/s \citep{BS71,Cy93}.
These particular values are not likely to be relevant for
plasma parcels in the solar wind, but we find it interesting to
note the parallels with this other application.
\item
For astrophysical ram-pressure stripping---which efficiently
removes mass from regions with gas pressure less than the
large-scale flow's ram pressure---it may be appropriate to assume that
$V_{\rm eff} \approx c_s$, the adiabatic sound speed of the gas.
For the solar corona, this hints at values around 150 km~s$^{-1}$.
\item
Another way to estimate $V_{\rm eff}$ is to examine hydrodynamic
models of large interstellar clouds that are destroyed when they
encounter shocks from nearby supernovae \citep{CM77,Kn94}.
\citet{Kn94} estimated that an initially spherical parcel would be
mostly destroyed over a cloud-crushing timescale
\begin{equation}
  \tau_{cc} \, \approx \, \sqrt{\frac{\rho_i}{\rho_e}}
  \frac{L}{|u_i - u_e|}
\end{equation}
where we have converted the original expressions into the notation
of this paper.
However, another way to estimate the timescale over which a parcel
undergoes substantial mass loss would be
\begin{equation}
  \tau_{m\ell} \, = \, \frac{M_i}{|dM_i / dt|}
  \, = \, \left( \frac{\rho_i}{\rho_e} \right)
  \frac{L V_{\rm eff}^2}{|u_i - u_e|^3}
\end{equation}
using Equation (\ref{eq:dMiAblate}).
Setting $\tau_{cc} = \tau_{m\ell}$, we can solve for the ablation
efficiency parameter
\begin{equation}
  V_{\rm eff} \, \approx \, |u_i - u_e|
  \left( \frac{\rho_i}{\rho_e} \right)^{-1/4}
\end{equation}
which is never far from just $|u_i - u_e|$ itself;
i.e., a few hundred km~s$^{-1}$ for our coronal parcels.
\end{enumerate}

\subsection{Approximate Analytic Solutions}
\label{sec:ablate:analytic}

Following Sections \ref{sec:drag:analytic} and \ref{sec:snowplow:analytic},
we describe a simplified system in which $u_e$, $\rho_e$, $A$, and $L$
are assumed to be constants.
Using the same dependent variables as above, there are two coupled
differential equations,
\begin{equation}
  \frac{dU}{dt} \, = \, - \frac{C_{\rm D} \, U^2}{QL}
  \,\,\,\,\,\,\,\,\, \mbox{and}
  \,\,\,\,\,\,\,\,\,
  \frac{dQ}{dt} \, = \, -\frac{U^3}{L V_{\rm eff}^2} \,\, .
  \label{eq:anablate}
\end{equation}
Inspired by the mathematical technique described by \citet{Br83},
we divide the $dQ/dt$ equation by the $dU/dt$ equation.
This gives a first-order separable equation for $Q$ as a
function of $U$, which integrates to
\begin{equation}
  Q \, = \, Q_0 \exp \left(
  \frac{U^2 - U_0^2}{2 C_{\rm D} V_{\rm eff}^2} \right) \,\, .
  \label{eq:QUablate}
\end{equation}
As time increases, $U$ decreases, so $Q$ also decreases from its
initial value of $Q_0$ to a final asymptotic value
\begin{equation}
  Q_{\infty} \, = \, Q_0 \exp \left(
  \frac{- U_0^2}{2 C_{\rm D} V_{\rm eff}^2} \right) \,\, .
\end{equation}
However, the above solution may not apply for all time.
Under the present set of approximations, when $Q$ first becomes
equal to 3/2, it means that $M_i = M_e$ and the Heaviside step
function in Equation (\ref{eq:dMiAblate}) becomes equal to zero.
Thus, once $Q$ reaches a value of 3/2, it remains fixed at this
value for all future times.
In that case, $U(t)$ behaves like the mass-conserving model
(Equation (\ref{eq:Umasscon})) during this later phase of parcel
evolution.

During the times when $M_i > M_e$ and ablation occurs, 
Equation (\ref{eq:QUablate}) can be inserted back into
Equation (\ref{eq:anablate}) and integrated again.
There does not appear to be a closed-form solution for $U(t)$,
but an inverted solution for $t(U)$ can be written.
If we define a new dimensionless variable
$y^2 = U^2 / (2 C_{\rm D} V_{\rm eff}^2)$, then
\begin{equation}
  \Omega t
  \, = \, \sqrt{\pi} \left[ \mbox{erfi} (y_0) - \mbox{erfi} (y)
  \right] + \frac{e^{y^2}}{y} - \frac{e^{y_0^2}}{y_0}
  \label{eq:erfisol}
\end{equation}
where
\begin{equation}
  \Omega \, = \,
  \frac{2^{1/2} C_{\rm D}^{3/2} V_{\rm eff}}{Q_{\infty} L}
\end{equation}
and the imaginary error function
$\mbox{erfi} (z) = -i \, \mbox{erf} (iz)$
is real-valued for a real argument $z$.
In the weak-ablation limit of $V_{\rm eff} \rightarrow \infty$,
Equation (\ref{eq:erfisol}) reduces to Equation (\ref{eq:Umasscon})
as it should.

\subsection{Numerical Solutions}
\label{sec:ablate:numerical}

As in Sections \ref{sec:drag:numerical} and \ref{sec:snowplow:numerical},
a large set of numerical models was constructed to determine if
any combinations of parameters could reproduce the observed deceleration
trend of \citet{Df14}.
In this case, similar ranges of random values for $C_{\rm D}$,
$\sigma$, and $L_0$ were explored.
The initial parcel density ratio $(\rho_i / \rho_e)_0$ was sampled
from a smaller range of 1.001 to $10^{+3}$; i.e., excluding values
less than unity that would not exhibit any ablation.
The efficiency parameter $V_{\rm eff}$ was sampled randomly
from a uniform grid (in logarithm) over twelve orders of magnitude:
from $10^{-6} c_s$ to $10^{+6} c_s$, with a fiducial value of
the coronal sound speed $c_s = 150$ km~s$^{-1}$.

Unfortunately, there were no combinations of input parameters
that agreed with the observed deceleration trend.
The ``best'' models were ones with extremely weak ablation
(i.e., the largest values of $V_{\rm eff}$), and these were
virtually identical to the mass-conserving models discussed in
Sections \ref{sec:drag:analytic}--\ref{sec:drag:numerical}.
Minimum $\chi^2$ values were similarly around 6, for minimum radii
of about 6.5~$R_{\odot}$ and typical values of
$\omega_0 \approx 1.7 \times 10^{-6}$ s$^{-1}$.
Many of the ablated solutions ended up with overdensity ratios
$\rho_i / \rho_e \lesssim 1$ for $r_i$ between 7 and 12 $R_{\odot}$.
Parcels of this kind may not exhibit enough scattered-light contrast
to ever be visible with instruments like COR2, so it is probably
not surprising that no parcels with these kinematic properties were
observed.

\section{Observational Constraints on Initial Conditions}
\label{sec:obscon}

The snowplow model of Section \ref{sec:snowplow} appears
to be able to reproduce the observed deceleration trend of inflowing
parcels in the extended corona.
However, we have not addressed the origin of the assumed initial
condition of $u_{i,0} = -83.6$ km~s$^{-1}$ at
$r_{i,0} = 12.4 \, R_{\odot}$.
For the 11 days of COR2 observations analyzed by \citet{Df14},
there are several possibilities:
(1) that $r_{i,0}$ was consistently below the Alfv\'{e}n radius, in
which case it is possible for the parcels to be moving at either
Alfv\'{e}nic or sub-Alfv\'{e}nic speeds in the frame of the solar wind,
(2) that $r_{i,0}$ remained above the Alfv\'{e}n radius, in which case
inward-moving parcels must have been ``shot out of a cannon''
at a locally supra-Alfv\'{e}nic speed,
(3) that the Alfv\'{e}n radius drifted in radius over those 11 days
so that both possibilities (1) and (2) applied over some of that time,
or (4) the solar wind exhibited multiple Alfv\'{e}n radii during this time.
Nontrivial situations like (3) and (4) are discussed further in
Section \ref{sec:conc}.

In the remainder of this section, we use in~situ data to estimate
an expected range of downward flow speeds for coronal perturbations
at 12.4~$R_{\odot}$.
First, it is necessary to estimate the value of $r_{\rm A}$ itself
for a given set of heliospheric conditions.
There have been several proposed empirical methods for using data at
1~AU to estimate the radial distance of the Alfv\'{e}n surface
\citep[see, e.g.,][]{Ks10,ZH10,Go14,TC16,KK19,Liu21}.
In steady-state, one can combine the equations of mass and
magnetic flux conservation to see that the quantity
$\rho u^2 / V_{{\rm A}r}^2$ should remain constant along a magnetic
field line.
We refer to Equation (\ref{eq:VA}) to note that the Alfv\'{e}n
speed $V_{{\rm A}r}$ is defined here using only the radial component
of the magnetic field $B_r$,
and we now use $u$ to refer to the large-scale wind speed
external to the parcel, rather than $u_e$ as above.
Thus, because $u^2 / V_{{\rm A}r}^2 = 1$ at the Alfv\'{e}n radius,
we see that
\begin{equation}
  \frac{\rho_{\rm A}}{\rho_{\rm E}} \, = \,
  \left( \frac{u}{V_{{\rm A}r}} \right)^2_{\rm E}
  \label{eq:rhorat}
\end{equation}
where subscript A refers to the Alfv\'{e}n radius and subscript
E refers to the Earth at 1~AU.
Obtaining a measurement of the right-hand side of
Equation (\ref{eq:rhorat}) is straightforward, and it constrains the
ratio of the density at the Alfv\'{e}n radius to the density at 1~AU.

The density ratio described above can be converted into a heliocentric
radial distance by examining models of solar-wind acceleration.
We took a large database of one-dimensional ZEPHYR models---i.e.,
30 models from \citet{CvB07} and 289 models from \citet{CvB13}---and
produced distributions of the radii corresponding to specific
values of $\rho_{\rm A}/\rho_{\rm E}$.
The models from \citet{CvB07} were created with idealized magnetic
fields corresponding to coronal holes, helmet streamers, and the
open-field parts of active regions.
The models from \citet{CvB13} were created by performing a potential-field
extrapolation from a near-equatorial patch of quiet Sun observed by
the the Synoptic Optical Long-term Investigations of the Sun (SOLIS)
facility in 2003.

\begin{figure}[!t]
\epsscale{1.15}
\plotone{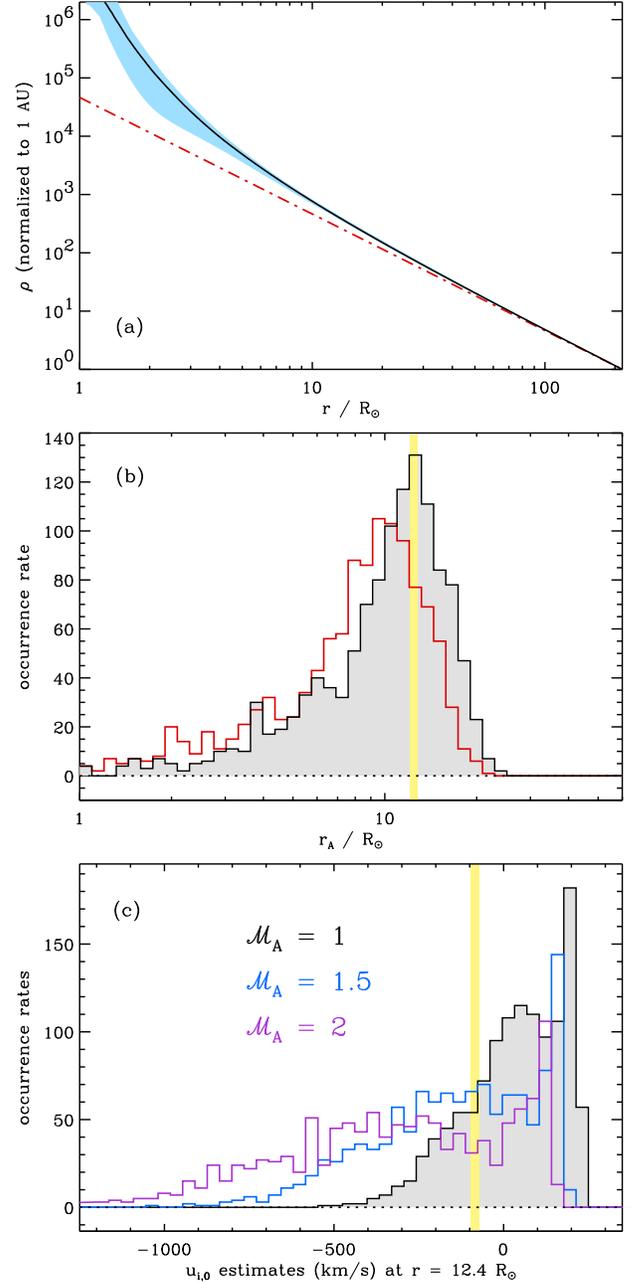}
\caption{(a) Ratio of density $\rho$ at various radii
to its value at 1~AU, computed from a constant wind-speed model
(red dot-dashed curve), from the median of a set of ZEPHYR models
(black solid curve), and from the full range of ZEPHYR models
(light-blue region).
(b) Histogram of values for $r_{\rm A}$ computed from OMNI fast-wind
data, in combination with the median ZEPHYR model (black curve)
and the constant wind-speed model (red curve).
Initial parcel radius $r_{i,0}$ is shown as a yellow bar.
(c) Histogram of inward-characteristic speeds computed from
Equation (\ref{eq:Cminus}) using Alfv\'{e}n Mach numbers
${\cal M}_{\rm A} = 1$ (black curve),
1.5 (blue curve), and 2 (purple curve), compared with the
observed initial parcel speed $u_{i,0}$ (yellow bar).
\label{fig08}}
\end{figure}

Figure \ref{fig08}(a) shows the radial dependence of the median values
of derived distributions of $\rho_{\rm A}/\rho_{\rm E}$,
as well as the minimum and maximum radii corresponding to each value
of the density ratio.
For example, if Equation (\ref{eq:rhorat}) indicates that a field-line
has a value of $\rho_{\rm A}/\rho_{\rm E} = 10^3$, one can infer from
the ZEPHYR models that this occurs at a median radius
$r_{\rm A} = 8.9 \, R_{\odot}$, with a relatively small spread around
that value of 8.4--9.3 $R_{\odot}$.

\citet{TC16} proposed a more straightforward approximation for finding
a direct solution of Equation (\ref{eq:rhorat}).
If the Alfv\'{e}n radius is large enough that negligible solar-wind
acceleration occurs above it, then one can assume $u =$~constant.
Far enough from the Sun, this implies $\rho \propto r^{-2}$, so
the left-hand side of Equation (\ref{eq:rhorat}) can be replaced
simply by $(r_{\rm E}/r_{\rm A})^2$,
where $r_{\rm E} = 1$~AU, and the expression can be solved analytically
for $r_{\rm A}$.
Figure \ref{fig08}(a) shows how this constant wind-speed
model provides a slightly smaller value of $r_{\rm A}$, for any
given measured density ratio $\rho_{\rm A}/\rho_{\rm E}$, than
do the ZEPHYR models that include solar-wind acceleration.

Ideally, we would prefer to carry out these measurements for the
specific solar-wind streams connected to the polar coronal holes
observed by \citet{Df14} in August 2007.
However, it is not clear whether these regions were magnetically
connected to any spacecraft in the ecliptic plane or elsewhere
in the solar system (e.g., Ulysses).
Thus, for the present analysis, we chose to examine a full solar cycle's
worth of OMNI data at 1~AU (calendar years 2008--2018) in order to
sample as many high-speed wind regions as possible that may have
been connected to large coronal holes.

One-minute-cadence OMNI data were used as a starting point
\citep[see, e.g.,][]{KP05}, then we extracted mean values of
$u$ and $V_{{\rm A}r}$ in successive two-hour bins.
With perfect data, that would provide 48,216 samples over the
selected 11-year period.
However, there were frequent data gaps (which were identified easily)
and CMEs (which were identified using the criteria of
\citeauthor{XB15} \citeyear{XB15}).
In order to isolate the data that accurately sample the ambient
solar wind, we rejected any bin containing more than 40 one-minute
data points (out of 120) with bad data or CMEs.
This reduced the total sample to 23,172 bins.
We then kept only high-speed wind data (appropriate for large
coronal holes) with radial speeds between 600 and 900 km~s$^{-1}$,
which reduced the sample further to only 1,208 bins.

Figure \ref{fig08}(b) shows the result of converting the measured
distribution of density ratios to values of $r_{\rm A}$.
When using the median ZEPHYR density trend, the median value of
$r_{\rm A}$ was 10.9~$R_{\odot}$, which is close to the value
of 10.73~$R_{\odot}$ corresponding to the baseline coronal-hole
model shown in Figure \ref{fig04}.
For the distribution of ZEPHYR-derived results for $r_{\rm A}$,
64\% of the values fall below the initial parcel radius
$r_{i,0} = 12.4 \, R_{\odot}$.
When using the \citet{TC16} density trend, the median value of
$r_{\rm A}$ was 8.7~$R_{\odot}$, and 81\% of the data correspond to
$r_{\rm A} < 12.4 \, R_{\odot}$.

Because $r_{\rm A}$ is sometimes larger than 12.4~$R_{\odot}$
and sometimes it is smaller, it is necessary to extrapolate from
$r_{\rm A}$ {\em both up and down} in order to estimate
the initial speeds of parcels.
Our goal is to first compute the linear characteristic speeds
of radial Alfv\'{e}nic perturbations ($C_{\pm} = u \pm V_{{\rm A}r}$)
at values of $r_{i,0}$ that may be above or below $r_{\rm A}$.
The OMNI measurement of radial wind speed 1~AU can be used in
combination with mass-flux conservation to estimate the value of
$u = V_{{\rm A}r}$ at the Alfv\'{e}n radius,
\begin{equation}
  u_{\rm A} \, \approx \, u_{\rm E}
  \left( \frac{\rho_{\rm E}}{\rho_{\rm A}} \right)
  \left( \frac{r_{\rm E}}{r_{\rm A}} \right)^2
  \,\, ,
\end{equation}
where we assume that there is no superradial expansion between
$r_{\rm A}$ and 1~AU (i.e., $A \propto r^2$).
For the fast-wind OMNI data, the median value obtained with the
ZEPHYR density ratios was $u_{\rm A} = 403$ km~s$^{-1}$, with a
large standard deviation of about 100 km~s$^{-1}$ around that value.
For comparison, the ZEPHYR model shown in Figure \ref{fig04}
had $u_{\rm A} = 509$ km~s$^{-1}$.

In order to extrapolate up and down in the neighborhood of the
Alfv\'{e}n radius, we can approximate
$A \propto r^{p}$ and $\rho \propto r^{-q}$.
In Section \ref{sec:drag:eqns}, values of $p \approx 2.15$ and
$q \approx 2.45$ were provided for the baseline coronal-hole model
between 7 and 12 $R_{\odot}$.
Thus, mass-flux conservation and the definition of the Alfv\'{e}n
speed give
\begin{equation}
  u \, \propto \, r^{q-p}
  \,\,\,\,\,\,\,\,\, \mbox{and}
  \,\,\,\,\,\,\,\,\,
  V_{{\rm A}r} \, \propto \, r^{(q-2p)/2}  \,\, .
\end{equation}
These relations allow us to estimate the characteristic speeds
of parcels that exist at radii other than $r_{\rm A}$.
Here we focus solely on the downward mode, and we allow for
the possibility of nonlinear propagation.
Thus, we can write $C_{-} = u - {\cal M}_{\rm A} V_{{\rm A}r}$,
where an Alfv\'{e}nic Mach number ${\cal M}_{\rm A} = 1$ indicates
linear wave motion and ${\cal M}_{\rm A} > 1$ indicates
supra-Alfv\'{e}nic (possibly shock-driven) motion, and
\begin{equation}
  C_{-}(r) \, = \, u_{\rm A} \left[ 
  \left( \frac{r}{r_{\rm A}} \right)^{q-p}
  - {\cal M}_{\rm A} \left( \frac{r}{r_{\rm A}} \right)^{(q-2p)/2}
  \right] \, .
  \label{eq:Cminus}
\end{equation}
Figure \ref{fig08}(c) shows the resulting distributions of $C_{-}$
speeds computed at a radial distance of $r_{i,0} = 12.4 \, R_{\odot}$,
using the ZEPHYR-derived results for $r_{\rm A}$ and three choices
for ${\cal M}_{\rm A}$.

For the linear case (${\cal M}_{\rm A}=1$), the median value of the
downward characteristic speed is positive (i.e., $+53$ km~s$^{-1}$)
since this situation corresponds to the initial parcel radius
$r_{i,0} = 12.4 \, R_{\odot}$ sitting above the median Alfv\'{e}n
radius $r_{\rm A} \approx 10.9 \, R_{\odot}$.
For the baseline distribution shown in Figure \ref{fig08}(c), downward
characteristic speeds of $C_{-} \leq -83.6$ km~s$^{-1}$ occur only
in 21\% of the samples.
However, Equation (\ref{eq:Cminus}) shows that it is possible to shift
the distribution of $C_{-}$ to smaller (more negative) values by
increasing ${\cal M}_{\rm A}$.
In fact, the measured downward speed of $-83.6$ km~s$^{-1}$
can be obtained as the median of the distribution
when ${\cal M}_{\rm A}$ is increased to a value of roughly 1.4.
The trend for the values of $C_{-}$ to decrease as
${\cal M}_{\rm A}$ is increased is captured by a linear fit of
these results,
\begin{equation}
  \langle C_{-} (r_{i,0}) \rangle \, \approx \,
  (413 - 355 {\cal M}_{\rm A} ) \,\, \mbox{km} \, \mbox{s}^{-1}
\end{equation}
where the angle-brackets denote median values of the distributions
constructed for each choice of ${\cal M}_{\rm A}$.
Note that only a slightly nonlinear Mach number
(${\cal M}_{\rm A} \approx 1.17$) is needed to produce a median
downward characteristic speed of zero at 12.4~$R_{\odot}$.

Is it possible that the initial downward speeds of the parcels
measured by \citet{Df14} are actually supra-Alfv\'{e}nic in the
frame of the accelerating solar wind?
There are models of collisionless magnetic reconnection in which
the exhausts are affected by kinetic process and accelerated to
speeds higher than the local Alfv\'{e}n speed
\citep[e.g.,][]{Sh11,La13,Lu14}.
However, in cases where the Hall effect is responsible for
these rapid flows, the relevant spatial scales may be far smaller
than those of the observed parcels.
Models of jets formed by reconnection in the low corona often exhibit
either slow-mode or fast-mode MHD shocks \citep{YS96,Ya13,Ro18},
but in some cases the absolute speeds may be lower than
$V_{{\rm A}r}$ due to guide-field reconnection ignoring much of
the radial field.

The recent discovery of sharp magnetic ``switchbacks''
in the inner heliosphere by Parker Solar Probe may point to
the ubiquitous presence of highly nonlinear MHD structures in the
solar-wind acceleration region \citep{Bale19,Kasper19,Dudok20,Horb20}.
In fact, the list of proposed explanations for switchbacks
sounds very similar to the list of origin scenarios for inward-propagating
coronal parcels.
Phenomena such as magnetic reconnection, shear-driven instabilities,
and large-amplitude turbulent eddies have all been suggested
\citep[e.g.,][]{Sq20,Tn20,Rf20,Zn20,Shoda21}.
It may be the case that the same events produce both outward and inward
disturbances.

\section{Discussion and Conclusions}
\label{sec:conc}

Inspired by observations of decelerating inflows at heliocentric
distances between 7 and 12~$R_{\odot}$, we constructed dynamical
models of discrete plasma parcels in the corona.
We found that parcels with constant or decreasing mass are not
able to reproduce the pattern of deceleration observed by
\citet{Df14}.
However, parcels with increasing mass---i.e., undergoing
snowplow-like interactions with the surrounding solar-wind
plasma---can reproduce the observed deceleration pattern
\citep[see also][]{Tn16}.
We also found that the most likely initial conditions for these
parcels at $\sim$12~$R_{\odot}$ involve mildly nonlinear (i.e.,
supra-Alfv\'{e}nic) speeds like those associated with shocks or jets.

Although the models developed in this paper help point us to 
the most important physical processes at work, it will be useful to
supplement them with more realistic simulations of parcels associated
with, e.g., magnetic reconnection, Kelvin-Helmholtz instabilities,
or MHD turbulence.
For example, \citet{Ly20} simulated inflows at 2--6~$R_{\odot}$ from
intermittent magnetic reconnection at the tips of helmet streamers.
Self-consistent simulations extended to larger distances may tell us
why only specific values of the drag coefficient $C_{\rm D}$ and snowplow
efficiency $C_{\rm S}$ seemed applicable to the observed flow patterns.
Also, some of our assumptions about geometric inertia effects
(i.e., $C_{\rm A}$) and even the quadratic nature of the drag
may need to be revised in the light of new simulations
\cite[see, e.g.,][]{MG10,Ve20}.

Despite this paper's focus on hydrodynamic and (mostly ideal) MHD
processes, it may be necessary to include additional physics in models
of coronal inflows.
The presence of strong electron heat conduction leads to the existence
of collisionless {\em thermal fronts} that propagate at speeds of order
$c_s$ and $V_{\rm A}$ but depend on the local plasma properties in
different ways \citep[see, e.g.,][]{Br79,Ru85,Kr15}.
The physics of these conduction fronts may govern the observed
kinematics of polar jets seen above the limb with Hinode \citep{Sj07}
and in the coronal-hole magnetic network by IRIS \citep{Tian14}.
These jets may also be related to the ubiquitous polar plumes, which are
often seen to survive out to the Alfv\'{e}n surface
\citep{Df97,Rao08,Rao16}.

It has been proposed that the Alfv\'{e}n radius is more accurately
described as being a frothy ``Alfv\'{e}n zone,'' and that at any one
time there may be multiple points along a radial line at which
the wind speed equals the Alfv\'{e}n speed.
\citet{Df18} discussed how plasma parcels of different sizes may flow
at different speeds in such an environment because the characteristic
speeds can vary from one end of the parcel to the other.
In fact, there may be places where short-wavelength parcels propagate
inward and long-wavelength parcels propagate outward.
Also, with a complex enough three-dimensional structure, individual parcels
may undergo random-walk-like deflections---alternately in and out---over
their lifetimes.
Thus, the steadily decelerating inflow pattern detected with STEREO/COR2
may be the result of intermittently sampling a distribution rather
than tracing a laminar trajectory.
In 2023, the Polarimeter to Unify the Corona and Heliosphere (PUNCH)
will begin mapping low-contrast motions between 6 and 180 $R_{\odot}$
\citep{Df20}.
This, along with other future coronagraphs and heliospheric imagers,
will provide much better constraints about the dynamics and statistics
of inhomogeneous flows in the solar wind.

\acknowledgments

The authors gratefully acknowledge
Anna Tenerani, Yuhong Fan, and the PUNCH science team
for many valuable discussions.
The authors are also grateful to the anonymous referee for
many constructive suggestions that have improved this paper.
SRC's contribution to this work
was supported by the National Aeronautics and Space
Administration (NASA) under grants {NNX\-15\-AW33G} and
{NNX\-16\-AG87G}, and by the National Science Foundation (NSF)
under grant 1613207.
The National Center for Atmospheric Reseach is a major facility
sponsored by the NSF under Cooperative Agreement No.\  1852977.
This research made extensive use of NASA's Astrophysics Data System (ADS).
The authors acknowledge use of OMNI data from NASA's Space Physics
Data Facility OMNIWeb service.

\end{document}